\def\year{2021}\relax
\documentclass[letterpaper]{article} 
\usepackage{aaai21}  
\usepackage{times}  
\usepackage{helvet} 
\usepackage{courier}  
\usepackage[hyphens]{url}  
\usepackage{graphicx} 
\usepackage{natbib}  
\usepackage{caption} 
\usepackage[switch]{lineno}
\usepackage[table,xcdraw]{xcolor}

\usepackage{subfig}
\usepackage{algorithm}
\usepackage[algo2e,ruled]{algorithm2e}
\usepackage{color}

\usepackage{soul}
\usepackage{url}
\usepackage{courier}
\usepackage[utf8]{inputenc}
\usepackage{caption}
\usepackage{graphicx}
\usepackage{amsmath}
\usepackage{booktabs}
\usepackage{calrsfs}
\DeclareMathAlphabet{\pazocal}{OMS}{zplm}{m}{n}

\newcommand{\Lb}{\pazocal{L}}
\usepackage{booktabs}
\usepackage{amsmath}
\usepackage{amssymb,amsthm,enumitem}
\usepackage{pifont}
\usepackage{soul}
\usepackage{algorithm}
\usepackage[noend]{algpseudocode}

\newcommand{\xmark}{\ding{55}}%
\usepackage{pdfrender}
\newcommand*{\boldcheckmark}{%
  \textpdfrender{
    TextRenderingMode=FillStroke,
    LineWidth=.5pt, 
  }{\checkmark}%
}

\title{Self-supervised transfer learning of physiological representations \\ from free-living wearable data }

\author{%
  Dimitris Spathis$^1$, Ignacio Perez-Pozuelo$^{2,3}$, Soren Brage$^2$,\\ \textbf{Nicholas J. Wareham}$^2$, and \textbf{Cecilia Mascolo}$^1$  \\
  }
  
\affiliations{

 $^1$Department of Computer Science and Technology, University of Cambridge, UK\\
  $^2$MRC Epidemiology Unit, School of Clinical Medicine, University of Cambridge, UK\\
  $^3$Alan Turing Institute, UK\\
}

\begin{document}

\maketitle

\begin{abstract}

Wearable devices such as smartwatches are becoming increasingly popular tools for objectively monitoring physical activity in free-living conditions. To date, research has primarily focused on the purely supervised task of human activity recognition, demonstrating limited success in inferring high-level health outcomes from low-level signals. Here, we present a novel \textit{self-supervised} representation learning method using activity and heart rate (HR) signals without semantic labels. With a deep neural network, we set HR responses as the \textit{supervisory signal} for the activity data, leveraging their underlying physiological relationship. In addition, we propose a custom quantile loss function that accounts for the long-tailed HR distribution present in the general population. We evaluate our model in the largest free-living combined-sensing dataset (comprising $>$280k hours of wrist accelerometer \& wearable ECG data). Our contributions are two-fold: i) the pre-training task creates a model that can accurately forecast HR based only on cheap activity sensors, and ii) we leverage the information captured through this task by proposing a simple method to aggregate the learnt latent representations (embeddings) from the window-level to user-level. Notably, we show that the embeddings can generalize in various downstream tasks through transfer learning with linear classifiers, capturing physiologically meaningful, personalized information. For instance, they can be used to predict variables associated with individuals’ health, fitness and demographic characteristics, outperforming unsupervised autoencoders and common bio-markers. Overall, we propose the first multimodal self-supervised method for behavioral and physiological data with implications for large-scale health and lifestyle monitoring. 

\end{abstract}

\section{Introduction}
The advent of wearable technologies has given individuals the opportunity to unobtrusively track everyday behavior. Given the rapid growth in adoption of internet-enabled wearable devices, sensor time-series comprise a considerable amount of user-generated data~\cite{blalock2016extract}. However, extracting meaning from this data can be challenging, since sensors measure low-level signals (e.g., acceleration) as opposed to the more high-level events that are usually of interest (e.g., arrhythmia, infection or obesity onset). Most wearable devices, particularly those that are wrist-worn, incorporate accelerometry sensors, which are very affordable tools to objectively study physical activity patterns~\cite{doherty2017large}\footnote{Throughout this work, we refer to activity, movement and acceleration interchangeably as signals obtained from wearable accelerometers.}. However, since wearables are used in daily, unconstrained environments, activities like drinking coffee or alcohol, as well as stress, may confound simple heuristics. 


Deep learning models, on the other hand, do not require handcrafted feature engineering, capture the temporal dynamics of sequential data, and exploit the latent representations inherently present in this data~\cite{hammerla2016deep}. 
 These approaches have shown great promise in human activity recognition (HAR) tasks using wearable sensor data~\cite{yang2015deep,ma2019attnsense, alsheikh2015deep}, but rely on purely labeled datasets which are costly to collect ~\cite{bulling2014tutorial}. In addition, they are obtained in laboratory settings and hence might not generalize to free-living conditions where behaviours are more diverse, covering a long tail of activities~\cite{krishnan2018insights}.  

\begin{figure}[t!]
    \centering
    \includegraphics[width=1.\linewidth]{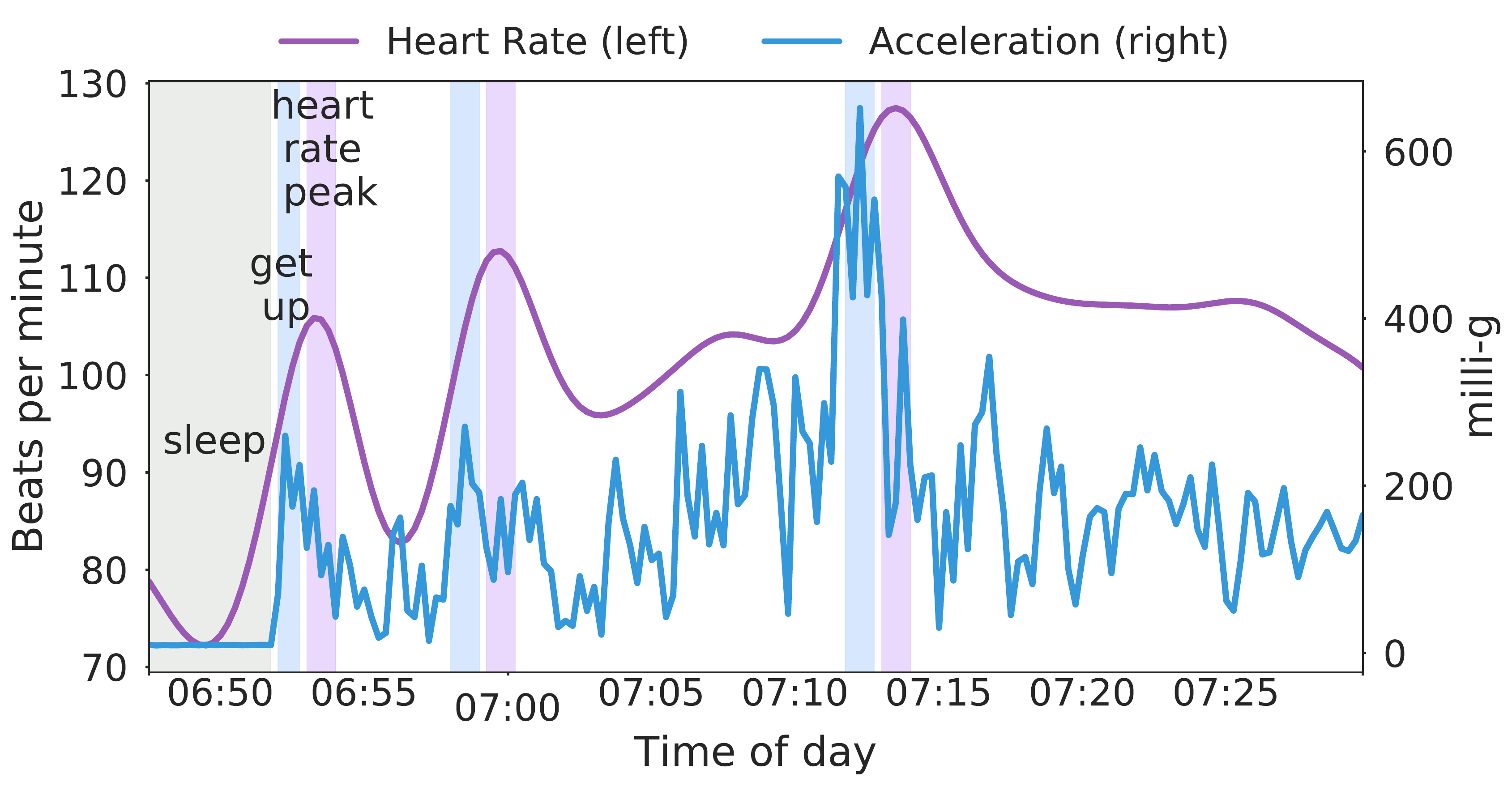}
	\caption{\textbf{Heart rate and acceleration temporal dynamics}. Illustrative visualization of the relationship between movement and heart rate responses (randomly selected participant). Shaded areas show this lagging relationship.}
	\vspace{-0.5cm}
    \label{fig:illustration}
\end{figure}

Unsupervised learning is a natural candidate to solve this label scarcity problem in wearable data, particularly given the vast amounts that can be collected in free-living conditions. Recent models have effectively utilized unlabeled activity data to learn useful summary representations of sensor signals~\cite{aggarwal2019adversarial}. Notwithstanding the value of these newly proposed methods, they only rely on a single stream of sensor data, usually movement data, and do not fully exploit the multimodal nature of modern wearable devices. Indeed, physical activity is characterized by \textit{both} movement and the associated cardiovascular response to movement (e.g., heart rate increases after exercise and the dynamics of this increase are dictated by fitness levels~\cite{jones2000effect}), thus, leveraging these two signals concurrently likely produces better representations than either signal taken in isolation. 
This relationship is conceptualized in Figure~\ref{fig:illustration}. Heart rate (HR) responses to exercise have been shown to be strongly predictive of cardiovascular disease (CVD), coronary heart disease (CHD) and all-cause mortality~\cite{savonen2006heart}. 

Multimodal learning has proven beneficial in supervised tasks such as fusing images with text to improve word embeddings~\cite{mao2016training}, video with audio for speech classification~\cite{ngiam2011multimodal}, or different sensor signals for HAR~\cite{radu2018multimodal}. However, all these approaches rely on the modalities being used as parallel inputs, limiting the scope of the resulting representations. Self-supervised training allows for mappings of aligned coupled data streams (e.g. audio to images~\cite{owens2016ambient} or, in our case, activity to heart rate), using unlabeled data with supervised objectives~\cite{lan2019albert}.



In this work, we present \textit{Step2Heart}, a general-purpose self-supervised feature extractor for wearable data, which leverages the multimodal nature of modern wearable devices to generate participant-specific representations. This architecture can be broken into two parts: 1) The new pre-training task forecasts ECG-level quality HR in real-time by only utilizing activity signals, 2) then, we leverage the learned representations of this model to predict personalized health-related outcomes through transfer learning with \textit{linear} classifiers. We hypothesize that this mapping captures more meaningful information than  autoencoders trained on activity data or other bio-markers.

\begin{figure*}
    \centering
    \includegraphics[width=0.99\linewidth]{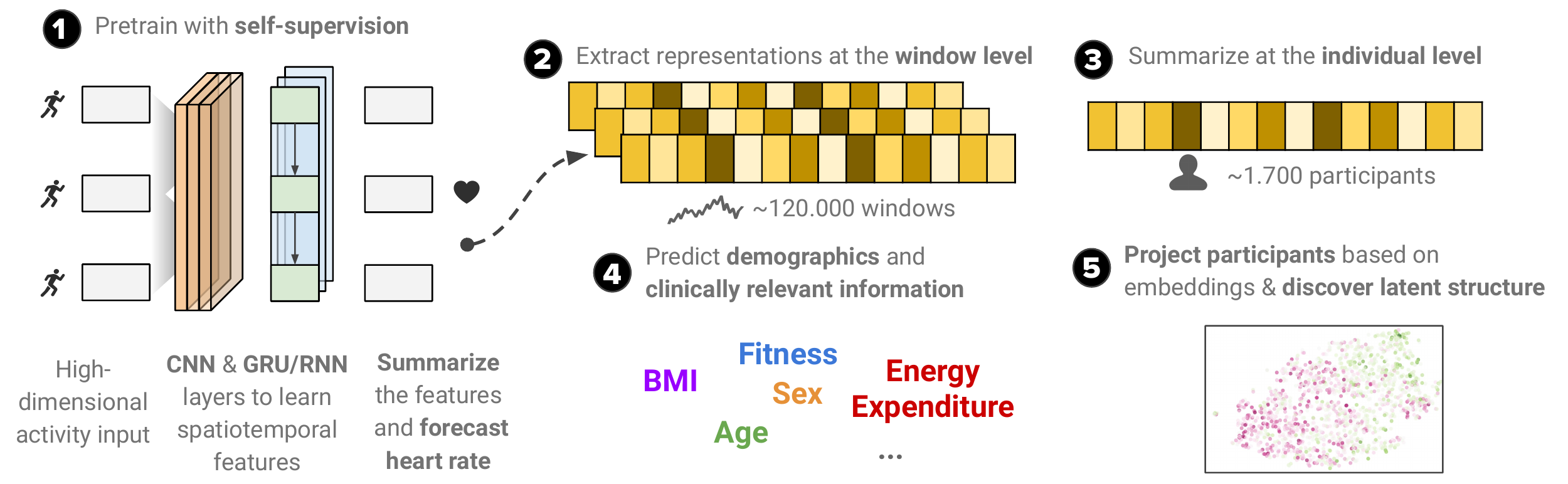}
	\caption{\textbf{Schematic of model architecture and tasks}. 
	}
	\vspace{-0.3cm}
    \label{fig:model_hero}
\end{figure*}

This paper puts forward four key technical contributions:

\begin{itemize}

 \item  We propose a novel \textit{self-supervised} model and a pre-training task which maps activity data to HR responses. Through this architecture, our model learns \textit{physiologically meaningful} user-level representations that can then be used for a variety of practical downstream tasks that are \textit{personalized} to the users' unique physiology.
 
  \item  For pre-training, we introduce a joint \textit{loss function} that acts as a regularizer to traditional MSE by using the quantiles of the predictive density of the model in order to approximate the long-tails of HR data observed in the real world. 

 \item  We evaluate this model in the largest multimodal wearable ECG and wrist accelerometry dataset, including over 1,700 participants tracked for a week, along with associated health outcomes measured with clinical lab equipment.  We perform ablation tests to show the performance of different modalities and components to the architecture. 

 \item  We perform a set of downstream, transfer learning tasks by aggregating the window-level features to user-level ones and showcase the value captured by the learned \textit{embeddings} through strong performance at inferring physiologically meaningful variables, outperforming autoencoders and common bio-markers. For example, our models achieve an AUC of 0.70 for Body Mass Index (BMI) prediction and an AUC of 0.80 for Physical Activity Energy Expenditure.

\end{itemize}


We envision our work having applications in facilitating the comprehensive monitoring of cardiovascular health and fitness at scale. Beyond health, this approach can be adapted to other parallel time-series tasks and their associated outcomes. The proposed model is summarized in Figure~\ref{fig:model_hero} while our code will be made publicly available upon acceptance.

\section{Related work}

\textbf{Objective monitoring of physical behaviors.} Large scale studies of physical activity leveraging mobile devices' built-in accelerometers have shown promise as global physical activity surveillance tools, demonstrating inequality across different countries and world regions~\cite{althoff2017large}. 
Mobile and wearable sensors allow for continuous and ubiquitous monitoring of an individual's physical activity profiles, which combined with cardio-respiratory information, provides valuable insights into that individuals' health and fitness status~\cite{mandsager2018association}. 
Hence, the possibility of measuring individuals' physiological characteristics in free-living conditions is of great interest for research, clinical and commercial applications.


\textbf{Machine learning for wearable sensing.} 
Recently, advances in deep learning architectures for sequential modeling based upon wearable and mobile sensing have been used for health predictions and recommendations~\cite{ballinger2018deepheart,schwab2019phonemd}. For example, \textit{FitRec}, an LSTM-based approach to modelling HR and activity data for personalized fitness recommendations was able to learn activity-specific contextual and personalized dynamics of individual user HR profiles during exercise segments~\cite{ni2019modeling}. This approach is helpful but requires prior segmentation of activities, which can be a constraint when applying these techniques in free-living, unconstrained conditions. Additionally, previous work has explored forecasting heart rate from movement data, however this was done on a much smaller scale (3 users) and used PPG sensors instead of the more accurate ECG as ground-truth~\cite{mcconville2018online}.

\textbf{Self-supervised pre-training.} Recent work using self-supervised learning has shown state-of-the-art results in computer vision~\cite{lee2017unsupervised,jenni2018self}, signal processing and natural language processing~\cite{lan2019albert}. Use cases in wearable and mobile sensing have been limited to human activity recognition using mobile devices~\cite{saeed2019multi} and emotion recognition using ECG data~\cite{sarkar2019self}, both using a single modality (acceleration or ECG), whereas we explore the unsupervised combination thereof guided by their physiological relationship. Our work is also inspired by the cardiovascular signature network introduced by Hallgrimmson et al~\cite{hallgrimsson2018learning}. However, this is an auto-encoder based approach requiring a historical input of 1-month of data for its prediction which renders the whole setup not feasible for real time applications. Furthermore, the data used is much more aggregated and limited in terms of outcomes than the data presented here. Overall, the generalizability of the learned embeddings is an under-explored area with some recent promising results in hospital operation room data~\cite{chen2020deep}, while abstract (non-sensor related) attributes like gender and age have been proved to be predictable with wearable embeddings~\cite{wu2020representation}. In sum, we believe that our work is the first multimodal general-purpose model to extract physiological and behavioral representations.

\section{Methods}

In this section, we provide a brief introduction to the problem formulation and notation used and then explore the model architecture and the associated methods proposed in this work.

\textbf{Problem formulation and notation.}  For this work, we assume $\mathcal{\textit{{N}}}$ samples, an input sequence $\mathcal{\textbf{X}}$ = (${\textbf{x}_{1}}$,...,${\textbf{x}_{N}}$) $\in \mathbb{R}^{N\times T\times F}$ and a target heart rate response  $\mathcal{\textbf{y}}$ = (${\textbf{y}_1}$,...,${\textbf{y}_{N}}$) $\in \mathbb{R}^{N}$. Additionally, we also consider contextual metadata like the hour of the day $\mathcal{\textbf{M}}$ = (${\textbf{m}_1}$,...,${\textbf{m}_{N}}$) $\in \mathbb{R}^{N\times F}$.
We use the same length $\mathcal{\textit{{T}}}$ for all sequences in our model. However, this sequence length is not a requirement and can be adapted based on the requirements of the task at hand or the granularity of the data. The intermediate representations of the model after training are $\mathcal{\textbf{E}}$ = (${\textbf{e}_1}$,...,${\textbf{e}_{N}}$) $\in \mathbb{R}^{N\times D}$ where $D$ is the latent dimension. These embeddings are aggregated at the user level $\mathcal{\tilde{\textbf{E}}}$ = ($\tilde{{\textbf{e}_1}}$,...,$\tilde{\textbf{e}_{N}}$) $\in \mathbb{R}^{\frac{N}{U} \times D}$, where $U$ is the number of users, in order to predict relevant outcome variables  $\mathcal{\tilde{\textbf{y}}}$ = ($\tilde{{\textbf{y}_1}}$,...,$\tilde{{\textbf{y}_{N}}}$) $\in \mathbb{R}^{N}$. Our full notation is summarized in Table \ref{tab:notation}. We employ two representation learning tasks: self-supervised pre-training and a downstream transfer learning task.

\noindent {\bf Upstream task: self-supervised pre-training and HR forecasting.} Given the accelerometer input sensor sequence $\mathcal{\textbf{X}}$ and associated metadata $\mathcal{\textbf{M}}$, predict the target HR $\mathcal{\textbf{y}}$ in the future. The input and target data shouldn't share temporal overlap in order to leverage the cardiovascular responses with the self-supervised paradigm by learning to predict the future. Motivated by population differences in heart rates, here we propose a custom \textit{quantile regression loss} to account for the tails of the distribution. This task by itself can be used for a reliable and real-time estimation of HR based on activity data.

\noindent {\bf Downstream task: transfer learning of learned physiological representations.} Given the internal representations $\mathcal{\textbf{E}}$ --usually at the penultimate layer of the aforementioned neural network \cite{sanchez2019machine}--, predict relevant variables $\mathcal{\tilde{{\textbf{y}}}}$ regarding the users' fitness and health using traditional classifiers (e.g. Logistic Regression). Inspired by the associations between word and document vectors in NLP \cite{le2014distributed}, we develop a simple aggregation method of sensor windows to the user level. This is a common issue in the literature \cite{chen2019developing}.

{\footnotesize
\begin{table}[]
\begin{tabular}{p{2.2cm}p{5.4cm}} 
\toprule
Notation & Description  \\ \midrule
$\mathcal{D}_{train}$, $\mathcal{D}_{test}$    & training and testing set for the forecasting task \\

 $\mathcal{\textbf{X}}$, $\in \mathbb{R}^{N\times T\times F}$      &     input sensor sequences        \\
 
  $\mathcal{\textbf{M}}$, $\in \mathbb{R}^{N\times F}$      &     input user metadata         \\
 
 $\mathcal{\textbf{y}}$, $\in \mathbb{R}^{N}$      &  target heart rate response           \\
 
    $\mathcal{\textit{{N}}}$      &    number of data points (samples)        \\

   $\mathcal{\textit{{T}}}$          &   length of input sequence             \\
   
   $\mathcal{\textit{{F}}}$        &  number of features (attributes)             \\ 
   
      $\mathcal{\textit{{U}}}$        &  number of users \\
   
   $\mathcal{\tilde{D}}_{train}$, $\mathcal{\tilde{D}}_{test}$ & training and testing set for the transfer learning task \\
   
     $\mathcal{{{\theta}}}$         &   parameters (weights) of a trained neural network         \\
   
   $\mathcal{\textit{{D}}}$         &   dimension of latent space embedding           \\
   
    $\mathcal{\textbf{E}}$, $\in \mathbb{R}^{N\times D}$      &        embeddings matrix learned from activity to heart rate mapping       \\
     $\mathcal{\tilde{\textbf{E}}}$, $\in \mathbb{R}^{\frac{N}{U} \times D}$      &   embeddings matrix learned like $\mathcal{\textbf{E}}$ (aggregated at the user level)           \\
      $\mathcal{\tilde{\textbf{y}}}$, $\in \mathbb{R}^\frac{N}{U}$      &  target variable for transfer learning (user level)  \\  
   
\bottomrule  
\end{tabular}
\caption{\textbf{Notation.}} 
\vspace{-0.5cm}
\label{tab:notation}
\end{table}
}

\begin{algorithm2e}
\SetAlgoLined
  \SetKwInOut{Input}{Input}
   \SetKwInOut{Output}{Output}
 \Input{$\mathcal{\textbf{X}}$ (activity),  $\mathcal{\textbf{M}}$ (metadata),  $\mathcal{\textbf{y}}$ (future HR)}
 \Output{ $\mathcal{\tilde{\textbf{E}}}$ (embeddings),  $\mathcal{\tilde{\textbf{y}}}$ (target outcome) }
 \While{neural network $\mathcal{\theta}$ not converged}{
  pass $\mathcal{\textbf{X}}$ through CNN/RNN layers\; 
  pass $\mathcal{\textbf{M}}$ through reLU layers\;
  concatenate outputs in  $\mathcal{\textbf{E}}$\;
  backpropagate $\mathcal{\textbf{y}}$ with joint loss $\Lb$ (eq. \ref{quantile_eq})\;
 }
 use trained network $\mathcal{\theta}$ to extract embeddings $\mathcal{\textbf{E}}$\;
 aggregate $\mathcal{\textbf{E}}$ to the user-level $\mathcal{\tilde{\textbf{E}}}$ with average pooling\;
  train a linear model to predict target variables $\mathcal{\tilde{\textbf{y}}}$\;
 \caption{\textit{Step2Heart} model pseudocode}
 \label{algorithm}
\end{algorithm2e}

\subsection{Model architecture}
As shown in Figure~\ref{fig:model_hero} we propose \textit{Step2Heart}, a general-purpose deep neural network for embedding wearable data. Its layers receive high-dimensional activity inputs along with associated metadata and learn spatio-temporal dynamics in order to accurately predict HR responses. It uses stacked convolutional (CNN) and recurrent (RNN) layers building upon architectures like \textit{DeepSense}~\cite{yao2017deepsense}, which have been proven state of art in mobile sensing. Here we present each component of the model. An overview of the overall method is given as a pseudocode in Algorithm \ref{algorithm}.

\subsubsection{Layers.} 
Given an input sequence $\mathcal{\textbf{X}}$ = (${\textbf{x}_{1}}$,...,${\textbf{x}_{N}}$), it passes through a stack of CNN layers that scan over the sequences with 1D windows and learn filters $f : \{0, ... , k-1\}$ $\in \mathbb{R}$.
Please note that the 1D window learns patterns across all the parallel features of the 3D input tensor $\mathcal{\textbf{X}}$. The learned filters of the CNNs are then fed into stacked RNNs. Specifically we employ a fast variant of RNNs known as Gated Recurrent Units (GRU) \cite{cho2014learning}. 
Then, the GRU output 
passes through a pooling layer that performs global element-wise averaging in order to summarize all the timesteps of the 3D tensor to a 2D matrix. If needed, the representation after the pooling operation can be concatenated with other features or metadata after passing through feed forward \textit{ReLu} layers. We also refer to this representation at the penultimate layer, $\mathbf{E}$, or \textit{embeddings} matrix. 
Last, the final layer is a feed forward neural network with a linear activation which is appropriate for continuous prediction tasks.

\subsubsection{Custom loss function.}

Heart rates vary across large populations. As such, some individuals may reach very low ($<$50 bpm, at rest/sleeping) or high ($>$180 bpm during vigorous exercise)~\cite{tanaka2001age} generating very long tails on the heart rate distribution. In traditional regression, the aim is to minimize the squared-error loss function or MSE  $\Lb_{MSE}(\mathbf{y,f}) = \frac{1}{N}\sum^{N}_{i = 1} \Lb(y_i - f(\mathbf{x}_i))^2$ to predict a single point estimate, similarly, quantile regressions aim to minimize the quantile loss in predicting a certain quantile. As such, the higher the quantile, the more the quantile loss function penalizes underestimates and the less it penalizes overestimates. 


The loss for an individual data point in quantile regression is defined by:

\vspace{-.4cm}
\begin{equation}
 \label{quantile_xi}
\Lb(\xi_i |\alpha ) = \left\{\begin{matrix}
\alpha \xi _i & $if$ \; \xi_i \geq 0, \\ 
 (\alpha - 1) \xi _i & $if$ \; \xi_i <  0. 
\end{matrix}\right.
\end{equation}

where $\alpha$ is the required quantile (between 0 and 1) and 

$\xi_i = y_i - f(x_i)$, where $f(x)$ is the predicted (quantile) model and $y$ is defined by the observed value for input $x$. A more compact version of Eq. \eqref{quantile_xi} can be formulated as $\Lb(\xi_i |\alpha) = max(\alpha \xi _i, (\alpha - 1) \xi _i)$ where $\xi \in \mathbb{R}$ is the residual.
As such, the average quantile loss over the whole dataset is:

\vspace{-.3cm}
\begin{equation}
    \Lb_Q(\mathbf{y,f}|\alpha) = \frac{1}{N}\sum^{N}_{i = 1} \Lb(y_i - f(\mathbf{x}_i)|\alpha ) 
\end{equation}

The quantile loss (or tilted/pinball loss in the literature) can be seen as \textit{tilted} version of the $l_1$ loss which estimates the unconditional median. Instead, if a prediction falls below a given quantile (e.g. $\alpha =0.10$), the residual is scaled (or tilted) by its probability $\alpha$. Thus, we can obtain the conditional quantile by minimizing the empirical $\Lb_Q$ loss. This formulation is inspired by similar loss functions applied to transportation problems \cite{rodrigues2018beyond} as well as reinforcement learning \cite{dabney2018distributional}.

In practice, we are interested in different quantile levels for the predicted probability distribution, not only one. Let $\{a \}^{J}_{j=1}$ be a set of $J$ quantiles (e.g. $0.05$, $0.10$, ..) we propose a joint loss function that leverages the $\Lb_{MSE}$ and  $\Lb_{Q}$ loss for an arbitrary number of quantiles: 

\vspace{-.3cm}

\begin{equation}
\begin{aligned}\label{quantile_eq}
    \Lb_{MSE+Q} & = \frac{1}{N}\sum^{N}_{i = 1} \bigg((y_i - f(\mathbf{x}_i))^2 \\
    & + \sum^{J}_{j=1} max \Big( \alpha_j(y_i - f(\mathbf{x}_i)^{(\alpha_j)}),\\
    & (\alpha_j-1) (y_i - f(\mathbf{x}_i)^{(\alpha_j)}) \Big) \bigg)
\end{aligned}   
\end{equation}

which can be seen as a sum of the MSE and the respective quantile losses, represented in one scalar. This scalar is used as the new backpropagation objective. In our context, very athletic or sedentary people can be considered as long-tail outliers and we want our models to account for it. Intuitively, the proposed loss can be seen as a combination of multiple objective functions where the second term acts as a regularizer for the MSE. During our experiments in next sections we apply different ablations of these terms to evaluate their impact.

\begin{table}
\centering
\begin{tabular}{>{\raggedright\arraybackslash}p{3.8cm}p{0.2cm}p{0.2cm}p{2.5cm}} 
\toprule
 Feature & Seq. & Inp. & Unit  \\ \midrule
 
 \textbf{Sensor} & & & \\
   \hspace{3mm}Acceleration & $\boldcheckmark$ & $\boldcheckmark$ & $m/s^2$ \\

 \hspace{3mm}Heart Rate & $\boldcheckmark$ & \xmark & Beats per Minute (BPM) \\
  
 \hspace{3mm}Timestamp & $\boldcheckmark$ &$\boldcheckmark$ & N/A \\ \midrule
 
 \textbf{Metadata} & & & \\
\hspace{3mm}UserID & \xmark & \xmark & N/A \\ 
          \hspace{3mm}Height & \xmark & \xmark & Meters \\ 
          \hspace{3mm}Weight & \xmark & \xmark & Kilograms \\ 
          \hspace{3mm}Sex & \xmark & \xmark & Male--Female \\
          \hspace{3mm}Resting HR & \xmark & $\diamond$ & BPM \\
          \hspace{3mm}$VO_{2max}$ & \xmark & \xmark & $mL/min $ $\cdot$ $kg$\\ \midrule
  
\textbf{Derived} & & & \\         
\hspace{3mm}Triaxial Acceleration & $\boldcheckmark$ &$\boldcheckmark$ & $m/s^2$ \\ 

 \hspace{3mm}ENMO & $\boldcheckmark$ &$\boldcheckmark$ & milli-\textit{g}\\  

 \hspace{3mm}VM-HPF & $\boldcheckmark$ &$\boldcheckmark$ &  milli-\textit{g} \\  

 \hspace{3mm}PAEE & \xmark & \xmark & $J/min$ $\cdot$ $kg$ \\ 

 \hspace{3mm}Body Mass Index (BMI) & \xmark & \xmark & $kg/m^2$ \\ 

 \hspace{3mm}Month, Hour &  \xmark & $\diamond$ &  $\cos$-$\sin$ transform \\ 
\bottomrule  
\end{tabular}
\caption{\textbf{Data description}. \textit{Seq.} denotes sequential measurements (timeseries), while \textit{Inp.} the inputs to the pre-training task. ($\diamond$ feature used in some models, see Results)}
\label{tab:features}
\vspace{-0.5cm}
\end{table}



\section{Evaluation}



\label{sec:dataset}

\textbf{Dataset.} The \textit{Fenland} study is a prospective cohort study that includes 12,435 men and women who are between the ages of 35-65
. After a baseline clinic visit, a subsample of 2,100 participants were asked to wear a combined heart rate and movement chest sensor 
and a wrist accelerometer
on their non-dominant wrist.
All participants provided written informed consent and the study was approved by the University of Cambridge Ethics Committee.

\subsubsection{Study protocol.}
The \textit{chest ECG} measured heart rate and uniaxial acceleration in 15-second intervals 
while the \textit{wrist device} recorded 60 Hz triaxial acceleration. 
Participants were told to wear both monitors continuously 24/7. 
During a lab visit, all participants performed a treadmill test that was used to inform their $VO_{2}max$ (maximum rate of oxygen consumption and a golden measure of fitness). Resting Heart Rate (RHR) was measured with the participant in a supine position using the \textit{chest ECG}. HR was recorded for 15 minutes and RHR was calculated as the mean heart rate measured during the last 3 minutes. These measurements were then used to calculate the Physical Activity Energy Expenditure (PAEE) \cite{brage2004branched}. 

\textbf{Pre-processing.} All participant heart rate data collected during free-living conditions underwent pre-processing for noise removal. 
Similarly, all  accelerometer data was auto-calibrated to local gravity, non-wear time was inferred and participants with less than 72 hours of wear were removed. Magnitude of acceleration was calculated through the \textit{Euclidean Norm Minus One} (ENMO) and the \textit{high-passed filtered vector magnitude (VM-HPF)} (expressed in milli-g/mg per sample). Both the accelerometry and ECG signals were summarized to a common time resolution of one observation per 15 seconds and no further processing to the original signals was applied. Since the time can have a big impact on physical activity (sleeping, commuting or even the season of the year), we encoded the sensor timestamps using \textit{cyclical temporal features} $T_f$ \cite{chakraborty2019advanced}. Here we encoded the month of the year and the hour of the day as $(x,y)$ coordinates on a circle:

\vspace{0.3cm}
\noindent\begin{minipage}{.50\linewidth}
\begin{equation}
    T_{f_1} = sin \Big(  \frac{2*\pi*t}{max(t)} \Big)
\end{equation}
\end{minipage}%
\noindent\begin{minipage}{.50\linewidth}
\begin{equation}
    T_{f_2} = cos \Big(\frac{2*\pi*t}{max(t)} \Big)
\end{equation}
\end{minipage}%
\vspace{0.3cm}

where $t$ is the relevant temporal feature (hour or month). The intuition behind this encoding is that the model will ''see" that e.g. 23:59 and 00:01 are 2 minutes apart (not 24 hours). 

\begin{table}
\centering
\resizebox{0.46\textwidth}{!}{
\begin{tabular}{p{2.6cm}p{2.0cm}p{1.8cm}p{1.7cm}} 
\toprule
 & MSE & RMSE & MAE   \\ \midrule
 
        $\textit{Step2Heart}_{A}$   &    $144.61$ ($0.62$)   & $12.02$ ($0.02$) & $9.23$ ($0.03$)\\
              
      $\textit{Step2Heart}_{A/T}$      &  $143.65$ ($0.28$)  & $11.98$ ($0.01$) & $9.21$ ($0.03$)  \\
               
   $\textit{Step2Heart}_{A/R}$     & $91.76$ ($0.12$)  & $9.57$ ($0.00$) & $6.92$ ($0.03$)  \\
            
  $ \textit{Step2Heart}_{A/R/T}$       &  $\mathbf{91.11}$ ($0.37$)  & $\mathbf{9.54}$ ($0.01$) & $\mathbf{6.88}$ ($0.02$)    \\
            \midrule  
            
    Baselines  &   &  &  \\
             \hspace{3mm} Global mean  & $250.99$  & $15.84$ & $12.46$    \\
             \hspace{3mm} User mean  & $186.05$  & $13.64$ & $10.40$    \\   
          \hspace{3mm} XGBoost$_{A}$  & $162.92$ ($0.20$)  & $12.76$ ($0.00$) & $9.83$ ($0.00$)  \\

\bottomrule  
\end{tabular}}
\caption{\textbf{Forecasting task results.} Ablation test to compare the HR forecasting error using different input modalities and baselines.}
\label{tab:forecasting}
\vspace{-0.3cm}
\end{table}

\label{sec:procedure}\textbf{Training procedure.} To create appropriate training batches for deep learning, we segmented the signals into fixed \textit{non-overlapping} windows of 512 timesteps, each one comprising 15-seconds and therefore yielding a window size of approximately 2 hours. We divided our dataset into training and test sets randomly using an 80-20$\%$ split with the training set then being further split into training and validation sets (90-10$\%$). We ensured that the test and train set had disjoint user groups. Further, we normalized the data by performing min-max scaling on all features described on Table \ref{tab:features} (sequence-wise for timeseries and column-wise for tabular ones) on the training set and applying it to the test set. During training, the target data (HR bpm) is not scaled and the forecast is 15" in the future after the last activity input.

The neural network was built through a stack of 2 CNN layers of 128 filters each, followed by 2 Bidirectional GRU stacked layers of 128 units each (resulting in 256 features due to bidirectional passes). When using extra inputs (RHR or timestamp derived features), a \textit{ReLu} MLP of dimensionality 128 was employed for each one and its outputs were concatenated with the GRU output. We trained using the Adam \cite{kingma2014adam} optimizer for 300 epochs or until the validation loss stopped improving for 5 consecutive epochs \footnote{hyper-parameter search was conducted with different layer numbers, unit sizes, learning rates and optimizers and we evaluated their impact on the validation set.}. The quantiles we used were [0.01, 0.05, 0.5, 0.95, 0.99] so that they equally cover extreme and central tendencies of the heart rate distribution. The XGBoost baseline's hyperparameters were found through 5-fold cross validation and were then applied to the test set. Likewise, in the transfer learning task, we followed the same procedure for Logistic Regression.

For the transfer learning task, we studied if the learned embeddings $\mathbf{E}$ can predict user variables ranging from demographics to fitness and health. Since a slightly lower number of users (1506) had sufficient fitness data obtained from the lab test visit
, we report only their results (the users remained in the same train/test splits  $\mathcal{\tilde{D}}_{train}$ / $\mathcal{\tilde{D}}_{test}$ as earlier). To create binary labels we calculated the 50\% percentile in each variable's distribution on the training set and assigned equally sized positive-negative classes. Therefore, even continuous outcomes such as BMI or age become binary targets for simplification purposes (the prediction is high/low BMI etc). The window-level embeddings were averaged with an element-wise mean pooling to produce user-level embeddings\footnote{we experimented with min, max and median pooling over embeddings but yielded consistently worse results across all variables.}. Then, to reduce overfitting, Principal Component Analysis (PCA) was performed on the training embeddings after standard scaling and the resulting projection was applied to the test set. We examined various cutoffs of explained variance for PCA, ranging from 90\% to 99.9\%. Intuitively, lower explained variance retained fewer components; in practice the number of components ranged from 10 to 160.

\begin{table}
\centering
\resizebox{0.40\textwidth}{!}{
\begin{tabular}{p{1.9cm}p{1.9cm}p{1.7cm}p{1.7cm}} 
\toprule
&  MSE & RMSE & MAE   \\ \midrule

  $\textit{Step2Heart}_{A/R/T}$    &  & &   \\
              $\Lb_{MSE}$  & $91.11$ ($0.37$)  & $9.54$ ($0.01$) & $6.88$ ($0.02$)    \\
             $\Lb_{MSE+Q}$  & $90.94$ ($1.12$)  & $9.53$ ($0.05$) & $6.90$ ($0.10$)   \\   
             $\Lb_{0.5*MSE+Q}$  & $\mathbf{90.27}$ ($0.53$) & $\mathbf{9.50}$ ($0.02$) & $6.81$ ($0.05$)     \\   
             $\Lb_{Q}$  & $92.0$ ($0.16$) & $9.59$ ($0.00$) & $\mathbf{6.75}$ ($0.02$)   \\ 
\bottomrule  
\end{tabular}}
\caption{\textbf{Loss function results.} Ablation test to compare the best performing model with regards to different loss functions. }
\vspace{-0.5cm}
\label{tab:loss_function}
\end{table}

        
          \label{fig:distributions}

\subsection{Baselines and metrics} For our baselines, we used naive lower bounds that require no models as well as modern ML approaches:

\begin{itemize}
    
    \item  \textbf{Global mean:} Predicts $y_{i}$ at each time step as the global HR mean of the training set. This is a naive baseline that assumes all users have the same HR anytime but provides a good lower bound for this longitudinal dataset.
   
    \item  \textbf{User mean:} \textit{Personalized} baseline obtained by predicting $y_{i}$ at each time step as the mean value for all the user's $\mathbf{X}$ in the training set. This is similar to the previous baseline but considers the entire heart rate range of each user over the study week.
  
    \item  \textbf{Convolutional Autoencoder:} A convolutional autoencoder which compresses the input data ($\mathbf{X}\rightarrow\mathbf{X}$) with a reconstruction loss. This unimodal baseline uses movement data only and is conceptually similar to \cite{aggarwal2019adversarial, saeed2019multi}. The intuition behind this is to assess whether \textit{Step2Heart} learns better representations due to learning a multimodal mapping of movement to heart rate ($\mathbf{X}\rightarrow\mathbf{y}$). To make a fair comparison, it has similar number of parameters to the self-supervised models and we use the bottleneck layer to extract embeddings (128 dimensions). This baseline is used only for the transfer learning experiments.
  
     \item  \textbf{Gradient Boosting (XGboost)}: gradient boosting machines are among the best performing ML methods~\cite{chen2016xgboost}. Since XGboost cannot work directly with timeseries, we extracted the following statistical features from the sensor windows: mean, std, max, min, percentiles (25\%, 50\%, 75\%) and the slope of a linear regression fit. The final feature vector consists of 80 features. 
\end{itemize}

Given the continuous nature of the forecasting task, we use standard metrics such as the Root Mean Squared Error (RMSE),
Mean Squared Error (MSE),
and Mean Absolute Error (MAE) 
for our evaluation. 
For the transfer learning task, the evaluation metric is the Area under curve (AUC).

\section{Results}
\label{sec:results}

\subsection{Forecasting}
 We consider different ablation tests for \textit{Step2Heart} as well as several baselines and report the average and standard deviation of 3 runs. For our ablation tests we consider the same model with different inputs: acceleration features only (A), with temporal features (A/T), with resting heart rate (A/R) and with both temporal features and resting heart rates (A/R/T).

\textbf{Impact of the Resting Heart Rate.} All results are summarized in Table~\ref{tab:forecasting}. \textit{Step2Heart} outperforms all baselines for this forecasting task and, when including temporal features and RHR  (\textit{Step2Heart}(A/R/T)), all performance metrics improve, resulting in an RMSE of 9.54. We note that the RMSE is probably the most interpretable metric since it directly translates to the error in HR beats per minute. Given the acceleration input, the addition of the RHR appears to be the most significant one, improving the RMSE by $\sim$ 2.5 and validating previous research that highlights RHR as a powerful bio-marker \cite{fox2007resting}. 

\textbf{Implicit personalization}. Interestingly, the baselines also reinforce the importance of personalized approaches as the user mean baseline vastly outperforms the global mean. Our models implicitly learn personalized patterns outperforming all baselines. Given the strong results of the embeddings in demographic prediction we present in the next section, we postulate that these models learn personalized features which would not be possible with other methods that --for example-- require user-specific layers and might not scale in large-scale datasets \cite{jaques2017predicting}.

\textbf{Impact of the joint loss.} When comparing different loss functions with the best performing model \textit{Step2Heart}(A/R/T), we see (Table~\ref{tab:loss_function}) that the proposed loss function better captures the long tails of HR. The lowest error, 9.5 RMSE, is achieved when weighting the MSE loss with the rest of the quantiles ($\Lb_{0.5*MSE+Q}$). Notably the pure quantile model achieves the best MAE of 6.75. We understand that a model optimized with the MSE loss would achieve better MSE score and a model including the 50\% quantile would optimize the MAE score. Thus, for this experiment we evaluate the impact of the losses \textit{across} all 3 metrics. In this case, the joint losses achieve the best results; the $\Lb_{Q}$ model may achieve the best MAE but predicably falls short in the other metrics. Given the overlapping standard deviations of the joint models  ($\Lb_{0.5*MSE+Q}$ and $\Lb_{MSE+Q}$) we consider both to be our best models, however we pick the former as the one with the lowest average error.





\begin{table}
\centering
\resizebox{0.48\textwidth}{!}{
\begin{tabular}{p{2cm}p{0.5cm}p{0.5cm}p{0.5cm}p{0.7cm}p{0.5cm}p{0.5cm}p{0.5cm}p{0.7cm}p{0.5cm}p{0.5cm}p{0.5cm}p{0.5cm}}
\toprule
Outcome & \multicolumn{12}{c}{AUC}                                                \\ \midrule
 & \multicolumn{4}{c}{{Conv. Autoencoder}}
        & \multicolumn{4}{|c|}{$\textit{Step2Heart}_{A/T}$}        & \multicolumn{4}{c}{$\textit{Step2Heart}_{A/R/T}$}  \\ \cmidrule(l){2-13} 
PCA\textbf{*}    & \multicolumn{1}{l}{$90\%$} & $95\%$ & $99\%$ & $99.9\%$ & $90\%$     & $95\%$    & $99\%$    & $99.9\%$ & $90\%$     & $95\%$    & $99\%$    & $99.9\%$      \\ \midrule

V$O_2 max$ & $52.6$  & $52.6$ & $59.6$ & $61.8$ & $58.6$ & $60$ & $63.9$ & $64.5$ & $\mathbf{68.3}$ & $67.8$& $68$& $68.2$     \\

PAEE & $69.6$  & $70.0$ & $70.2$ & $71.8$& $74.7$ & $74.7$ & $77.5$ & $76.8$ & $78.2$ & $79.2$& $\mathbf{80.6}$& $79.7$    \\

Height & $60.8$  & $60.3$ & $75.9$ & $79.4$& $66$ & $67.4$ & $77.4$ & $\mathbf{82.1}$ & $70.3$ & $74$& $80.5$& $81.3$   \\ 

Weight & $56.5$  & $56.2$ & $70.3$ & $72.1$& $65.7$ & $67.6$ & $75$ & $77.2$ & $69.9$ & $70.7$& $\mathbf{77.4}$& $76.9$   \\

Sex & $66.7$  & $67.0$ & $86.5$ & $89.7$& $72.3$ & $72.9$ & $87.1$ & $93.2$ & $76.2$ & $81.5$ & $91.1$ & $\mathbf{93.4}$   \\

Age & $46.2$  & $46.3$ & $53.9$ & $59.5$& $55.0$ & $61.7$ & $66.2$ & $66.9$ & $61.1$ & $63.8$ & $67.3$ & $\mathbf{67.6}$   \\

BMI & $51.6$  & $51.5$ & $60.1$ & $61.2$& $62.8$ & $63$ & $68.2$ & $67.6$ & $64.7$ & $66.1$ & $67.8$ & $\mathbf{69.4}$   \\

Resting HR & $49.1$  & $49.4$ & $55.8$ & $55.4$& $56.7$ & $56.6$ & $\mathbf{62.7}$ & $61.7$ & \multicolumn{4}{c}{N/A}   \\

\bottomrule  
\end{tabular}}
\caption{\textbf{Transfer learning results}. Performance of embeddings in predicting variables related to health, fitness and demographic factors. A random baseline yields an AUC of 50. All values are $\times$100 for better legibility. (\textbf{*}percentage of explained variance by compressing the dimensionality of embeddings with PCA)}
\vspace{-0.5cm}
\label{tab:transfer}
\end{table}

\subsection{Transfer learning}
For this set of results, we use the best-performing model as shown above ($\Lb_{0.5*MSE+Q}$), extract embeddings and train linear classifiers for different outcomes.

\textbf{Generalizing in downstream tasks.} Quantitatively, the embeddings achieved strong results in predicting variables like users' sex,  height, PAEE and weight (0.93, 0.82, 0.80 and 0.77 AUC respectively). Also, BMI, $VO_2max$ and age are moderately predictable (0.70 AUC). The pure acceleration model (A/T) moderately predicts Resting HR (0.62 AUC), but this does not apply to the (A/R/T) since it already includes the RHR as input. Generally, the A/R/T model outperforms the A/T model showing that using the RHR as input is helpful, as discussed in the previous sections. 

Our results validate previous studies like~\cite{hallgrimsson2018learning} with different and very aggregated data. As a simple baseline, we followed their idea of using the RHR as a single predictor and we could not surpass an AUC of 0.55 for BMI and age. Also, the autoencoder baseline, which learns to compress  the activity data, under-performs when compared to \textit{Step2Heart}$_{A/T}$, illustrating that the proposed task of mapping activity to HR captures the physiological state of the user, which translates to more generalizable embeddings. We note that both approaches operate only on activity data as inputs. This shows that the embeddings carry richer information than single biomarkers or modalities by leveraging the relationship between physical activity and heart rate responses. Obtaining these outcomes in large populations can be valuable for downstream health-related inferences which would normally be costly and burdensome (for example a $VO_2max$ test requires expensive laboratory treadmill equipment and respiration instruments). Additionally, PAEE has been strongly associated with lower risk of mortality in healthy older adults~\cite{manini2006daily}. Similarly, $VO_2max$ is prospectively associated with the incidence of type 2 diabetes~\cite{katzmarzyk2005metabolic}.

\textbf{Impact of the latent dimensionality size.} From the representation learning perspective, we observe considerable gain in accuracy in some variables when retaining more dimensions (PCA components). More specifically, Sex and Height improve in absolute around +0.20 in AUC. However, this behavior is not evident in other variables such as PAEE and $VO_2max$, which seem robust to any dimensionality reduction. This means that the demographic variables leverage a bigger dimensional spectrum of latent features than the fitness variables which can be predicted with a subsample of the features.  These findings could have great implications when deploying these models in mobile devices and deciding on model compression or distillation approaches~\cite{hinton2015distilling}.

\begin{figure}
    \centering
    \includegraphics[width=0.90\linewidth]{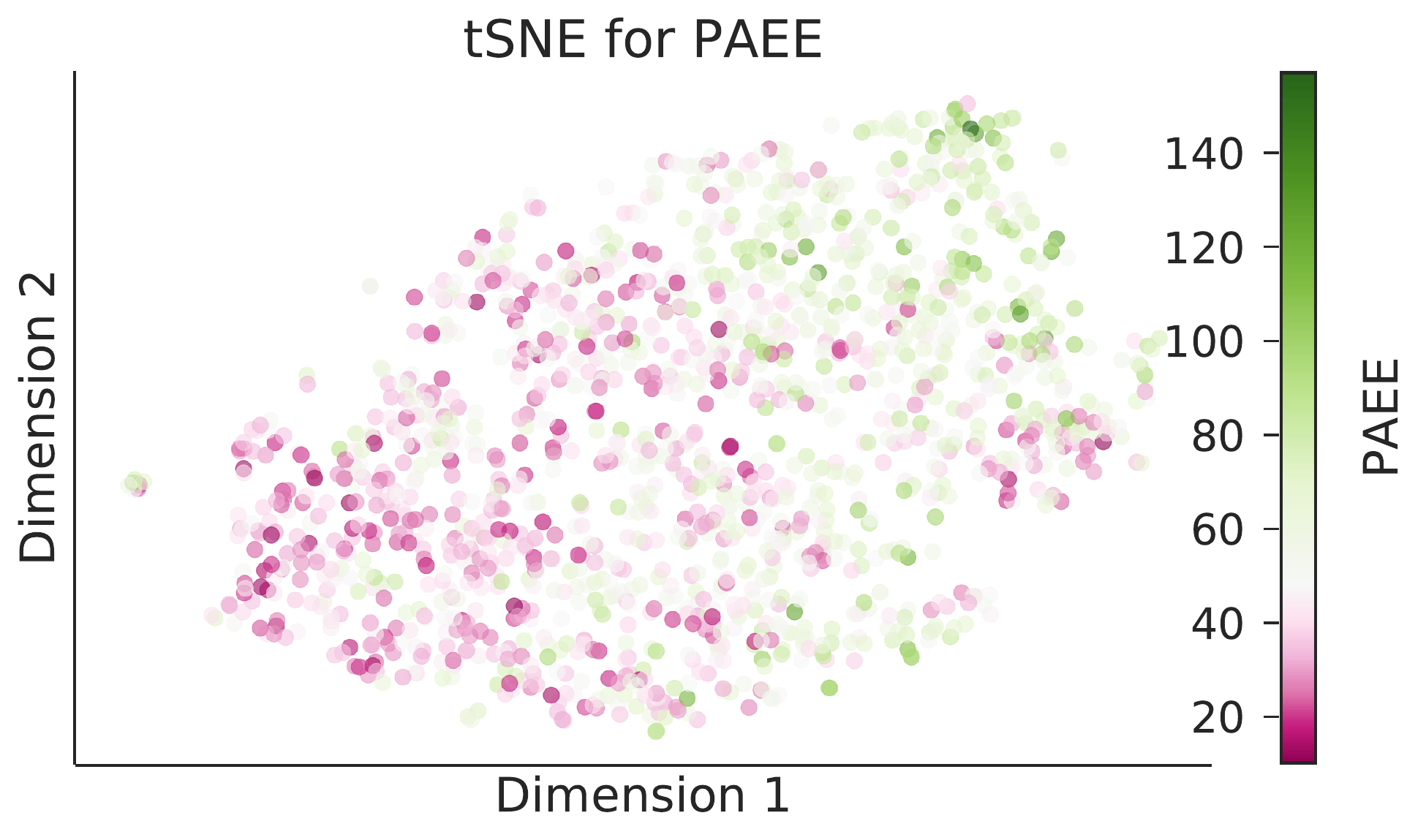}
	\caption{\textbf{Model embeddings for transfer learning visualized with t-SNE.} 2D representation of the embeddings for PAEE prediction. Color coding shows the extreme expenditures, since the median participant had PAEE 48 (white color). See Table \ref{tab:transfer} for full results.}
	\vspace{-0.5cm}
    \label{fig:tsne}
\end{figure}

\textbf{Visualizing the latent space.} Qualitatively, we visualized the resulting \textit{latent} space in 2D with t-Distributed Stochastic Neighbor Embedding (t-SNE)~\cite{maaten2008visualizing} as shown in Figure~\ref{fig:tsne}. In this setup, we used the embeddings of the entire dataset. We found that many of the outcomes, like the depicted PAEE cluster in their own specific regions. We color code the extreme PAEE users in order to illustrate that most normal users are grouped in the center but high/low PAEEs are diametrically opposed. 
These visualizations can help us understand common behaviours (similar users are neighbors in the latent space), would allow for risk stratification and potentially suggest interventions to specific groups (e.g. nutrition or exercise advice to high-risk BMI--obesity onset cluster).

\section{Conclusion}
Here we proposed a novel \textit{self-supervised} general-purpose feature extractor for wearable data. These features can be used for a variety of practical downstream tasks that are \textit{personalized} to the users' unique physiology. We evaluated this model with the largest dataset of its kind, including over 1,700 participants with combined heart and activity sensors for a week. Our model outperforms a set of strong baselines in both upstream and downstream tasks evaluated with ablation studies. 
 The task-agnostic embeddings achieved strong performance at inferring physiologically meaningful variables (BMI, fitness etc). By inspecting the embeddings we also noticed most outcomes improve with higher latent dimensionality, while some are invariant to its size. More fine-grained prediction of the outcomes is also left for future work. Last, this method proposed hereby could be applied to other domains where parallel time-series are prevalent (weather, traffic etc) in order to learn rich representations.

\newpage
\section{Ethical impact and broader implications}

The healthcare industry is undergoing an unprecedented digital transformation, producing and curating large amounts of data. Annotating all this data in order to feed to deep learning models for pattern recognition is impractical. Through self-supervised learning, we can leverage this unlabelled data to learn meaningful representations that can generalize in situations where ground truth is inadequate or simply infeasible to collect due to high costs. Such scenarios are of great importance in population health where we may be able to achieve clinical-grade health inferences with widely-adopted devices such as wearables and smartphones. Our work makes contributions in the area of transfer learning and subject-specific representations, one of utmost importance in machine learning for health.

Personalized health-representations like the ones arising from our models could raise some concerns if used maliciously for exclusionary insurance policies or unfair credit scoring, for example. However, we should clarify that our proposed model is a \textit{tool}, and like all tools might be subject to misuse. Hence, while the risks associated to \textit{Step2Heart} are minimal, it is paramount that future developments and use of this technology follow data governance principles that guarantee the rights of users, prevent misuse of data and promote trust in the rapidly evolving digital health ecosystem.

\section{Acknowledgements}
D.S was supported by the
Embiricos Trust Scholarship of Jesus College Cambridge, and EPSRC through Grant DTP (EP/N509620/1). I.P was supported by GlaxoSmithKline and EPSRC through an iCase fellowship (17100053). The authors declare that there is no conflict of interest regarding the publication of this work.


\bibliography{bibliography}
\end{document}